\documentclass[preprints,article,accept,moreauthors,pdftex]{Definitions/mdpi} 

\firstpage{1} 
\pubvolume{1}
\issuenum{1}
\articlenumber{0}
\pubyear{2024}
\copyrightyear{2020}
\datereceived{} 
\dateaccepted{} 
\datepublished{} 
\hreflink{https://doi.org/} % If needed use \linebreak
\Title{Constraints on Phase Transitions in Neutron Star Matter}

\TitleCitation{Constraints on Phase Transitions in Neutron Star Matter}

\Author{Len %MDPI: 1. Please carefully check the accuracy of names and affiliations. 2. Please do not delete our comments. 3. Please revise all questions that we proposed. And reply to them after the corresponding comment. Such as: “Author reply: It should be italic”; “Author reply: I confirm”; “Author reply: I have checked and revised all.” etc.
 Brandes * and Wolfram Weise *}

\AuthorNames{Len Brandes and Wolfram Weise}

\AuthorCitation{Brandes, L.; Weise, W.}
\address[1]{
 Physics %MDPI: Addresses should be arranged from small to large. We adjusted the address order, please confirm.
 Department, TUM School of Natural Sciences, Technical University of Munich, 85747 Garching, Germany}
\corres{\hangafter=1 \hangindent=1.05em \hspace{-0.82em}Correspondence: len.brandes@tum.de (L.B.); weise@tum.de (W.W.)}

\abstract{Recent inference results of the sound velocity in the cores of neutron stars are summarized.  Implications for the equation of state and the phase structure of highly compressed baryonic matter are discussed.  In view of the strong constraints imposed by the heaviest known pulsars,  the equation of state must be very stiff in order to ensure the stability of these extreme objects.  This required stiffness limits the possible appearance of phase transitions in neutron star cores.  For %MDPI: Please confirm if the bold is unnecessary and can be removed. The following highlights are the same.
 example,  a Bayes factor analysis quantifies strong evidence for squared sound velocities $c_s^2 > 0.1$ in the cores of 2.1 solar-mass and lighter neutron stars. % Please confirm that meaning has been retained 
  Only weak first-order phase transitions with a small phase coexistence density range $\Delta\rho/\rho < 0.2$ (at the 68\% level) in a Maxwell construction still turn out to be  possible within neutron stars.  The central baryon densities in even the heaviest neutron stars do not exceed five times the density of normal nuclear matter. In view of these data-based constraints,  much discussed issues such as the quest for a phase transition towards restored chiral symmetry  and the active degrees of freedom in cold and dense baryonic matter,  are reexamined.}


\begin{document}
%%%%%%%%%%%%%%%%%%%%%%%%%%%%%%%%%%%%%%%%%%
%\setcounter{section}{-1} %% Remove this when starting to work on the template.
%\section{How to Use this Template}

%The template details the sections that can be used in a manuscript. Note that the order and names of article sections may differ from the requirements of the journal (e.g., the positioning of the Materials and Methods section). Please check the instructions on the authors' page of the journal to verify the correct order and names. For any questions, please contact the editorial office of the journal or support@mdpi.com. For LaTeX-related questions please contact latex@mdpi.com.
%The order of the section titles is: Introduction, Materials and Methods, Results, Discussion, Conclusions for these journals: aerospace,algorithms,antibodies,antioxidants,atmosphere,axioms,biomedicines,carbon,crystals,designs,diagnostics,environments,fermentation,fluids,forests,fractalfract,informatics,information,inventions,jfmk,jrfm,lubricants,neonatalscreening,neuroglia,particles,pharmaceutics,polymers,processes,technologies,viruses,vision

\section{Introduction}

Within recent decades,  the observational data base for neutron stars has expanded
substantially.  The detailed analysis of this data has sharpened the empirical constraints on the equation of state (EoS) of matter deep inside the cores of these extreme objects~\cite{Lovato2022}.  In particular,  the existence of heavy neutron stars with masses around and above two solar masses implies that the EoS,  i.e.,  pressure $P(\varepsilon)$ as a function of energy density $\varepsilon$,  must be sufficiently stiff in order to support such massive compact stars against gravitational collapse.  Some previously discussed simple forms of exotic matter could, therefore, be excluded when their corresponding EoSs turned out to be too soft. 

The quest for a possible phase transition in cold and dense baryonic matter---from hadronic to quark degrees of freedom---has been a topic of prime interest for a long time~\cite{Fukushima2011, Aarts2023}.  In this context, a summary will be given of the present state of knowledge about the neutron star EoS as inferred directly from the existing observational data.  Of particular interest is the speed of sound in the cores of neutron stars.  Its behavior as a function of energy density is a sensitive indicator for phase transitions or continuous changes (crossovers) of degrees of freedom in a star's composition. 

Primary sources of information for the inference of the EoS are neutron star masses deduced from Shapiro delay measurements~\cite{Demorest2010,Antoniadis2013,Fonseca2016, Arzoumanian2018,Cromartie2020,Fonseca2021}, 
combined determinations of masses and radii inferred from X-ray data detected with the NICER telescope~\cite{Riley2019,Miller2019,Riley2021,Miller2021,Salmi2022}, 
and gravitational wave signals from binary neutron star mergers observed by the LIGO and Virgo Collaborations~\cite{Abbott2018,Abbott2019,Abbott2020}.
These data have served as basic input for a large variety of Bayesian and other inference analyses~\cite{Annala2020,Raaijmakers2021,Pang2021,Legred2021,Biswas2022, Ecker2022,Altiparmak2022, Huth2022, Lim2022, Annala2023,Somasundaram2023,Essick2023,Brandes2023, Brandes2023a,Mroczek2023, Han2023} in search of a range of data-compatible equations of state with controlled uncertainties.  Low-density constraints from nuclear physics around the equilibrium density of nuclear matter,  $\rho_0 = 0.16\,$fm$^{-3}$, are commonly introduced by referring to results of chiral effective field theory (ChEFT)~\cite{Drischler2022}.  At asymptotically high densities,  perturbative QCD (pQCD) is applicable and provides a constraint to be matched in extrapolations far beyond the conditions realized in neutron star cores~\cite{Gorda2021,Komoltsev2022}.   Bayesian inference studies have recently been further extended~\cite{Brandes2023a}  by the supplementary observational data reported in~\cite{Romani2022} %MDPI: We revised the reference citation order, please confirm.
% Author reply: I confirm.
 (the black widow pulsar PSR J0952--%MDPI: Please check if this should be a minus sign ("$-$" U+2212). Please unify the format in whole text. The same to following highlight without comment.
 % Author reply: This should be a dash '--', I have changed the appearance for the following cases.
0607 with a mass of $M = 2.35 \pm 0.17\,M_\odot$ in solar mass units).

This report presents a survey of results obtained in the detailed inference analysis of Ref.~\cite{Brandes2023a} and discusses possible interpretations.  A central topic is the evaluation of constraints and possible evidence for a first-order phase transition within the posterior uncertainties of the EoS deduced from the existing empirical neutron star data.  Later parts of this manuscript refer to ongoing considerations about the structure and composition of dense matter in neutron star cores.  
In view of the constraints emerging from the analysis of the observational data,  especially of the heaviest known neutron stars,  we discuss the quest for a chiral phase transition,  quark--hadron continuity and its realization as a soft crossover,  a Fermi liquid approach to neutron star matter,  and other related issues.

\section{Equation of State of Neutron Star Matter }
\label{sec-2}
\subsection{Observational Constraints}
\label{sec-2.1}
\subsubsection{Neutron Star Masses and Radii}
Information about the masses of the heaviest neutron stars derives primarily from precise Shapiro time delay measurements of pulsars orbiting in binary systems with white dwarf companions.  Three such massive objects, PSR J1614--2230~\cite{Demorest2010,Fonseca2016,Arzoumanian2018},  PSR J0348+0432~\cite{Antoniadis2013},
 and PSR J0740+6620~\cite{Cromartie2020,Fonseca2021},  have been established in the past:
\begin{eqnarray}
	&\text{PSR J1614--2230} \qquad &M = 1.908 \pm 0.016 \, M_\odot ~, ~ \label{eq:ShapiroMass1}\\
	&\text{PSR J0348+0432} \qquad &M = 2.01 \pm 0.04 \, M_\odot ~, \label{eq:ShapiroMass2}\\
	&\text{PSR J0740+6620} \qquad &M = 2.08 \pm 0.07 \, M_\odot ~. \label{eq:ShapiroMass3}
\end{eqnarray}
\textls[-24]{The %MDPI: Please confirm whether the paragraph below the formula needs to be indented. The same to following highlight without comment.
% Author reply: This paragraph does not need to be indented.
  heaviest neutron star observed so far was recently reported~\cite{Romani2022}---a black widow~pulsar,}
\begin{eqnarray}
	~~~\text{PSR J0952--0607}\qquad &M = 2.35\pm 0.17 \, M_\odot ~.~ \label{eq:BWMass}
	\end{eqnarray}
This result was obtained based on the detection of the black widow's companion star.  PSR J0952--0607 is also one of the fastest rotating pulsars.  With a spin period of 1.4 ms, it requires radius-dependent corrections for rotational effects, as described in~\cite{Brandes2023a}.  For example,  the rotating mass of a $2.35\,M_\odot$ star with an assumed radius of $R = 12$ km decreases by 3\% to an equivalent non-rotating mass of about 2.28 $M_\odot$.
 
Together with their masses the radii of neutron stars can be inferred from X-ray profiles of rotating hot-spot patterns measured with the NICER telescope.  Two neutron stars have been investigated in this way~\cite{Riley2019,Riley2021}: 
\begin{eqnarray}
	\text{PSR J0030+0451} &\qquad M = 1.34^{+0.15}_{-0.16} \, M_\odot ~, \qquad &R=12.71^{+1.14}_{-1.19}\,\text{km}~,\label{eq:NICER1}\\
         \text{PSR J0740+6620} &\qquad M = 2.072^{+0.067}_{-0.066} \, M_\odot ~, \qquad &R=12.39^{+1.30}_{-0.98}\,\text{km}~.\label{eq:NICER2}
\end{eqnarray}
All of the mass and radius data are included as inputs when setting constraints for the neutron star EoS.  In the case of PSR J0740+6620, the analysis of the NICER measurement includes the Shapiro delay result.  Therefore, only the data from NICER have  been used in the Bayesian inference dataset.  An alternative analysis of the NICER data was also  performed by a second independent group~\cite{Miller2019,Miller2021}.  Both results are mutually compatible within their~uncertainties.

\subsubsection{Binary Neutron Star Mergers and Tidal Deformabilities}
Gravitational wave signals produced by a merger of two neutron stars in a binary have been detected by the LIGO and Virgo collaborations~\cite{Abbott2019,Abbott2018,Abbott2020}.  These signals are interpreted using theoretical waveform models that depend on the mass ratio of the two neutron stars, $M_2/M_1$,  and a mass-weighted combination of their tidal deformabilities.  Based on information from the GW170817 event, the (dimensionless) tidal deformability for a 1.4 solar mass neutron star was deduced~\cite{Abbott2018}:
\begin{eqnarray}
\Lambda_{1.4} = 190^{+390}_{-120} ~.\label{eq:tidal1}
\end{eqnarray}
The GW170817 event was further investigated together with electromagnetic signals.  In one such analysis~\cite{Fasano2019}, the following masses and tidal deformabilities were reported for the individual neutron stars in the binary:
\begin{eqnarray}
	&\qquad M_1 = 1.46^{+0.13}_{-0.09} \, M_\odot ~, \qquad &\Lambda_1 =255^{+416}_{-171}~,\nonumber \\
         &\qquad M_2 = 1.26^{+0.09}_{-0.12} \, M_\odot  ~, \qquad &\Lambda_2=661^{+858}_{-375}~.\label{eq:tidal2}
\end{eqnarray}
The  GW190425 event, which could have also been a binary neutron star merger,  was also included in the inference analysis~\cite{Abbott2020}.

\subsection{Inference of Sound Velocity and the EoS in Neutron Stars}
\label{sec-2.2}

Based on the data in the previous subsection, the squared sound velocity and the equation of state 
\begin{eqnarray}
c_s^2(\varepsilon) = \frac{\partial P(\varepsilon)}{\partial\varepsilon}~~~\text{and}~~~P(\varepsilon)= \int_0^\varepsilon d\varepsilon'\,c_s^2(\varepsilon')~,
\label{eq:sound}
\end{eqnarray}
of neutron star matter can be inferred using Bayesian methods.  In our recent work~\cite{Brandes2023,Brandes2023a}, the ChEFT constraint at low baryon density was used in the Bayesian inference procedure as a likelihood ({\it not} as a prior) within a conservative window of baryon densities~\cite{Essick2020},  $\rho \lesssim 1.3\,\rho_0$.  The equations of state were interpolated to pQCD results at asymptotically high densities.  Unlike %MDPI: Please confirm if the bold is unnecessary and can be removed. The following highlights are the same.
% Author reply: I confirm the bold can be removed
 the conclusions drawn in~\cite{Komoltsev2022},  and as explained in detail in~\cite{Brandes2023a}, we did, however, not find a strongly constraining influence of pQCD extrapolations down to neutron star densities.  Possible reasons for this discrepancy were also discussed in~\cite{Komoltsev2023}.

\begin{figure}[H]
\includegraphics[width=8.8cm]{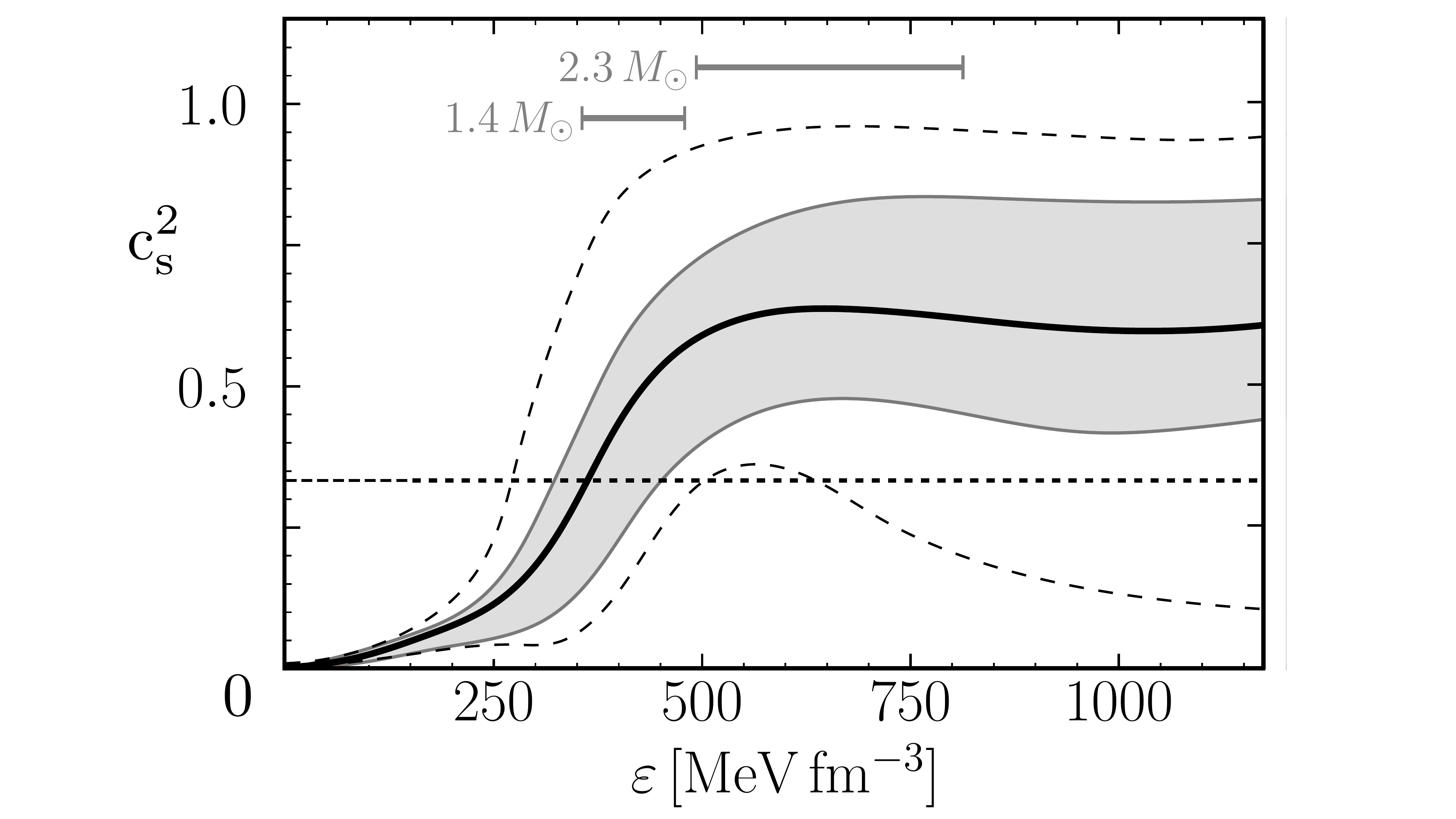}
\caption{Squared %MDPI: We moved the figure to the end of the first mention, please confirm the revision.
% Author reply: I confirm.
 speed of sound~\cite{Brandes2023a}
 as a function of the energy density inferred from the empirical dataset listed in Section \ref{sec-2.1}.  Median (solid curve) and highest posterior density credible bands % Please confirm that meaning has been retained 
 % Author reply: The meaning has not been retained, the correct terminus is 'highest posterio density credible bands'
  at levels of 68\% (gray band) and 95\% (dashed curves) are displayed.  Intervals are indicated for the 68\% ranges of central energy densities in the cores of 1.4 and 2.3 $M_\odot$ neutron stars. \label{fig1} }
\end{figure}

The result of the inferred sound velocity in Figure~\ref{fig1}
shows,  within the $68\%$ highest posterior density credible bands, % Please confirm that meaning has been retained 
% Author reply: The meaning has not been retained, the correct terminus is '68% highest posterior density credible bands'. I have changed the sentence accordingly.
  a rapid increase in $c_s^2$ beyond the conformal limit of 1/3 at energy densities relevant for a wide range of neutron stars with masses $M \sim$ 1.4--2.3 $M_\odot$.  A quantification in terms of a corresponding Bayes factor~\cite{Legred2021,Brandes2023,Brandes2023a} demonstrates  extreme evidence for $c_s^2$ exceeding the conformal bound inside neutron stars.

\begin{figure}[H]
\includegraphics[width=8.8cm]{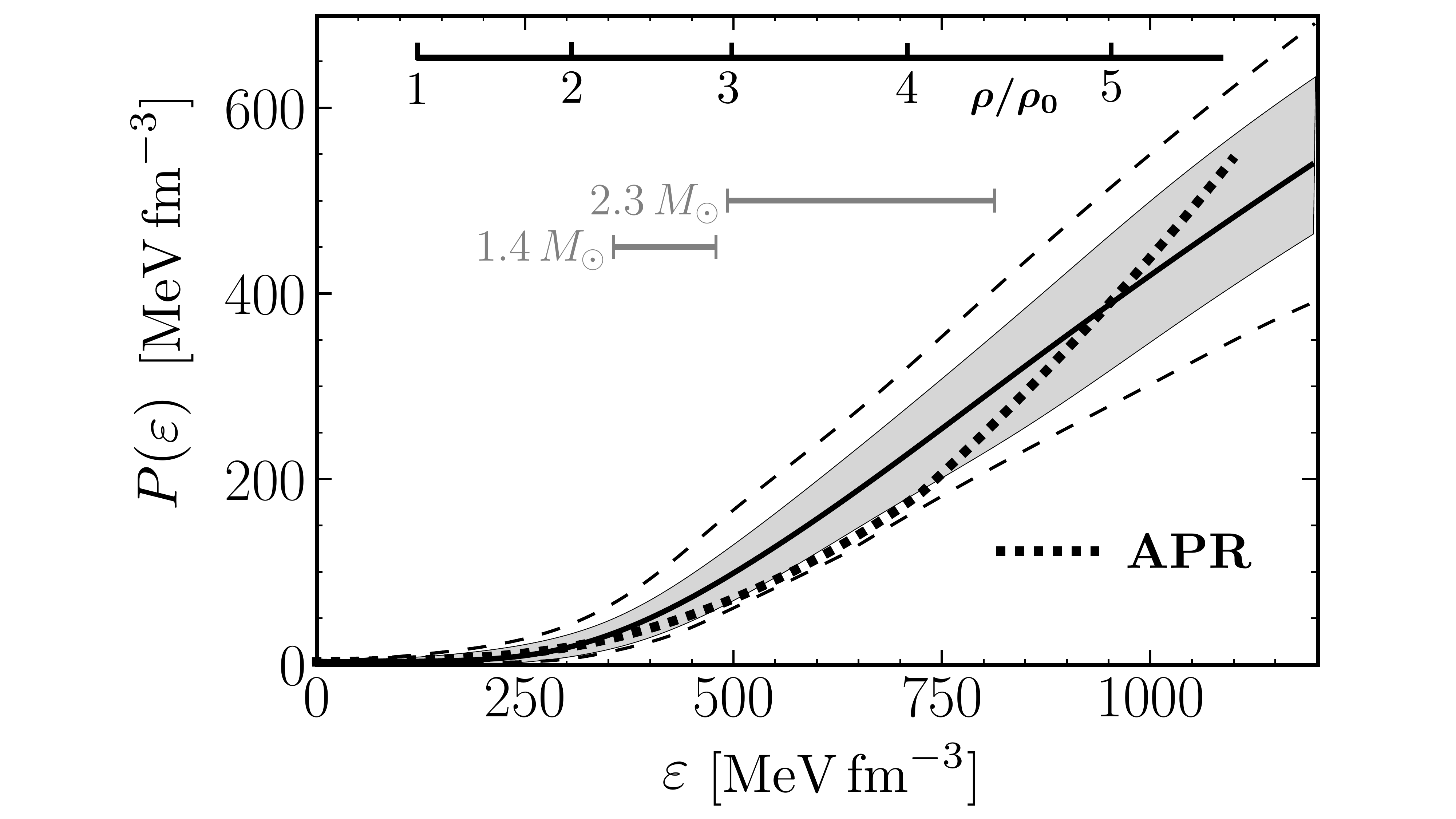}
\caption{Equation of state $P(\varepsilon)$~\cite{Brandes2023a} deduced 
 from the inferred sound velocity based on the empirical dataset listed in Section \ref{sec-2.1}.  The median,  68\%, and 95\% posterior credible bands are displayed, as in Figure~\ref{fig1}.  The APR EoS~\cite{APR1998} (dotted line) is shown for comparison.  Also shown for orientation is the baryon density scale, $\rho/\rho_0$ (in units of the equilibrium density of nuclear matter $\rho_0 = 0.16$ fm$^{-3}$),  which was computed for the median of $P(\varepsilon)$. \label{fig2} }
\end{figure}

\begin{figure}[H]
\includegraphics[width=8.8cm]{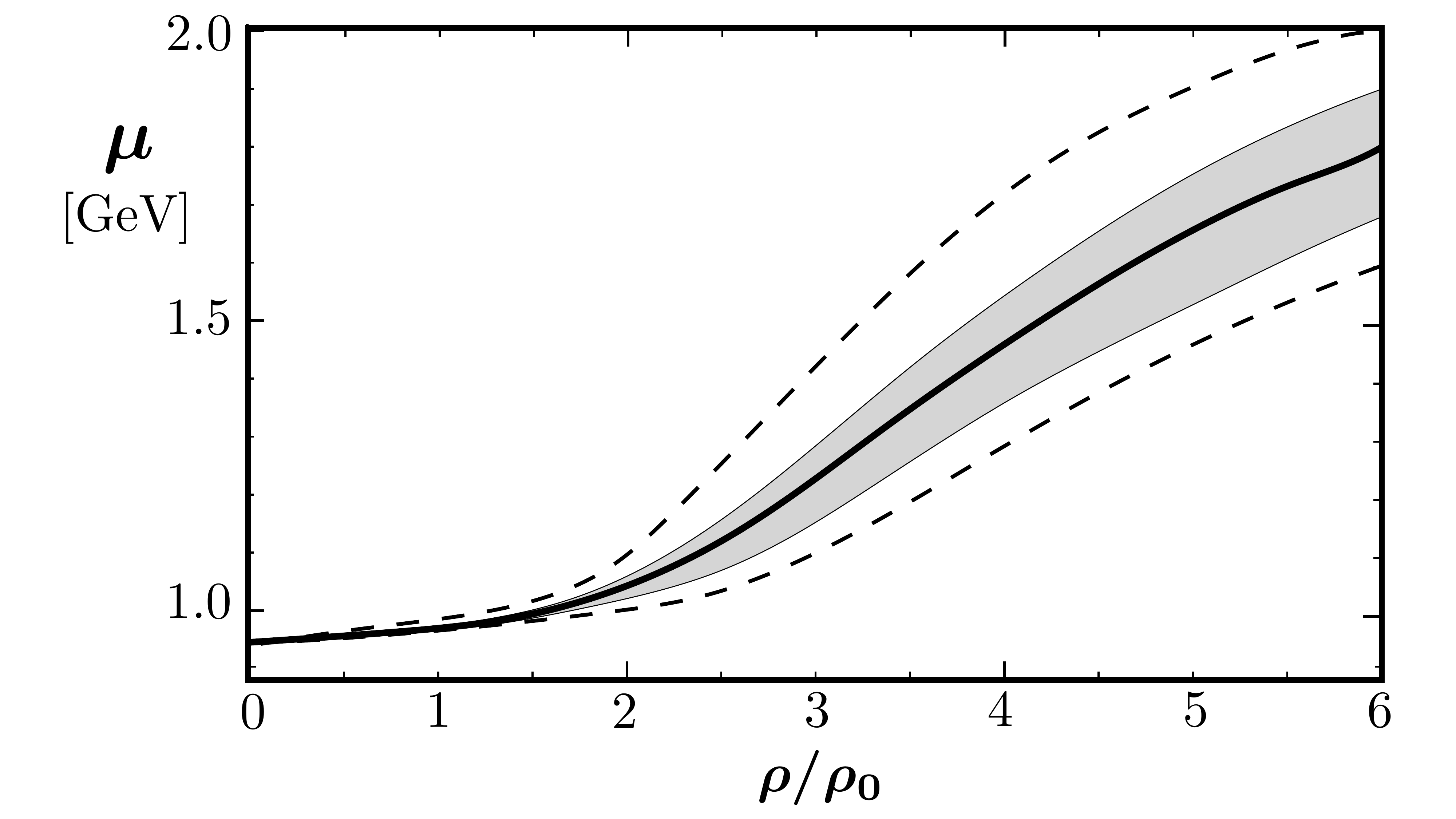}
\caption{ Baryon chemical potential %MDPI: Please check if the bold in figure should be retained, and keep consistent format in whole main text.
  % Author reply: Here the phase 'Baryon chemical potential' was altered to chemical potential of baryon, this is not the same! The bold in the figure can be removed to keep a consistent format throughout the text.
  $\mu(\rho)$ in neutron star matter,  normalized at $\mu(\rho=0) = 939$ MeV.  Posterior credible bands~\cite{Brandes2023a} at the 68\% level (gray band) and 95\% level (dashed lines) are shown.\label{fig3} }     
\end{figure}

Figure~\ref{fig2} presents the inferred posterior bands for the EoS.  Notably,  the additional inclusion of the heaviest  pulsar PSR J0952--0607 in the dataset leads to a more rapidly rising pressure as compared to previous inference results, which do not incorporate these data.  Within the range of energy densities 0.5--0.8 GeV/fm$^3$ reached in the core of a 2.3 solar mass star and corresponding to baryon densities between 3.0$\rho_0$ and 4.6$\rho_0$,  the pressure even exceeds that of the time-honored APR EoS~\cite{APR1998}.
A further quantity of interest is the chemical potential associated with the conserved baryon number,
\begin{eqnarray}
\mu(\rho) = \frac{\partial\varepsilon(\rho)}{\partial\rho} = \frac{P+\varepsilon}{\rho}~,
\end{eqnarray}
where the latter equality is the Gibbs--Duhem relation at a temperature of zero.  By solving the equation $\rho = \rho_l\exp\int_{\varepsilon_l}^{\varepsilon(\rho)}d\varepsilon'[\varepsilon' + P(\varepsilon')]^{-1}$ for $\varepsilon(\rho)$ with a boundary condition $\varepsilon_l = \varepsilon(\rho_l)$ chosen at a suitably low density $ \rho_l$, the baryon chemical potential can be computed for any given equation of state $P(\varepsilon)$ within the uncertainty bands in Figure~\ref{fig2}.  The resulting $\mu(\rho)$ bands in Figure~\ref{fig3} show a steep rise at baryon densities of $\rho > 2\,\rho_0$, indicating that strongly repulsive forces are at work.  The agnostic approach for the inference of the sound speed on which these results are based does not permit  distinction between different species of constituents inside neutron stars,  so $\mu$ represents the total chemical potential from all active degrees of freedom carrying baryon number: $\mu = \sum_i x_i\mu_i$,  where $x_i=\rho_i/\rho$ is the fraction corresponding to each baryonic species $i$.   However, irrespective of whether these species are nucleons,  other baryonic composites, or quarks,  the empirically inferred behavior of $\mu$ demonstrates that these degrees of freedom must be correlated by strongly repulsive forces as the density increases  in order to build up the necessary high pressure in the cores of neutron stars to support two solar masses and beyond.  We return to this discussion again at a later stage.

\begin{figure}[H]
\includegraphics[width=8.8cm]{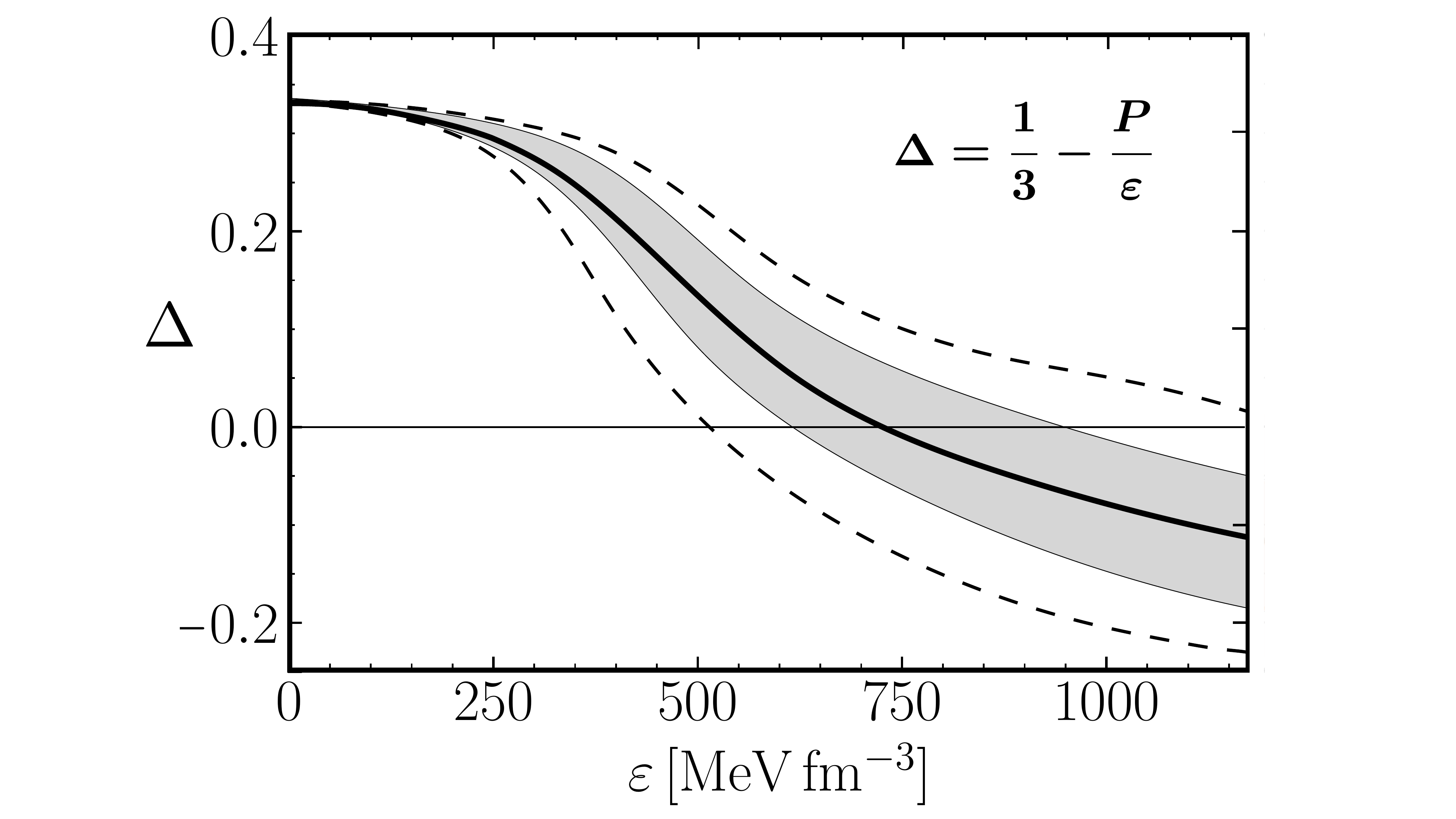}
\caption{Median and credible bands~\cite{Brandes2023a}
of the trace anomaly measure,  $\Delta = \frac{1}{3} - \frac{P}{\varepsilon}$,  at the 68\% level (gray band) and 95\% level (dashed lines).\label{fig4} }     
\end{figure}

The approach to conformality in strongly interacting matter at high baryon density has been an important theme in several recent investigations~\cite{Fujimoto2022, Marczenko2023,Rho2023,Ma2023}.   A key quantity is the trace anomaly measure,
\begin{eqnarray}
\Delta = \frac{g_{\mu\nu}T^{\mu\nu}}{3\varepsilon} = \frac{1}{3} - \frac{P(\varepsilon)}{\varepsilon}~.\label{eq:traceanomaly}
\end{eqnarray}
Conformal matter is characterized by an equation of state $P = \varepsilon/3$,  so the signature for its occurrence is $\Delta \rightarrow 0$ over an extended energy density range.  The posterior credible bands for the trace anomaly measure deduced from the data-based EoS are shown in Figure~\ref{fig4}.  Starting from $\Delta = 1/3$ at zero energy density,  the median of $\Delta$ turns negative at $\varepsilon\sim 0.7\,\text{GeV/fm}^3$,  entering a high-pressure domain with $P > \varepsilon/3$.  This crossing appears within the range of energy densities possibly realized in the cores of the heaviest neutron stars.  At much higher energy densities beyond those displayed in Figure~\ref{fig3},  pQCD then implies a switch back to positive $\Delta$ before the asymptotic limit $\Delta\rightarrow 0$ is approached.

\vspace{-6pt}

\subsection{Selected Neutron Star Properties}
\label{sec-2.3}

\begin{figure}[H]
\includegraphics[width=8.8cm]{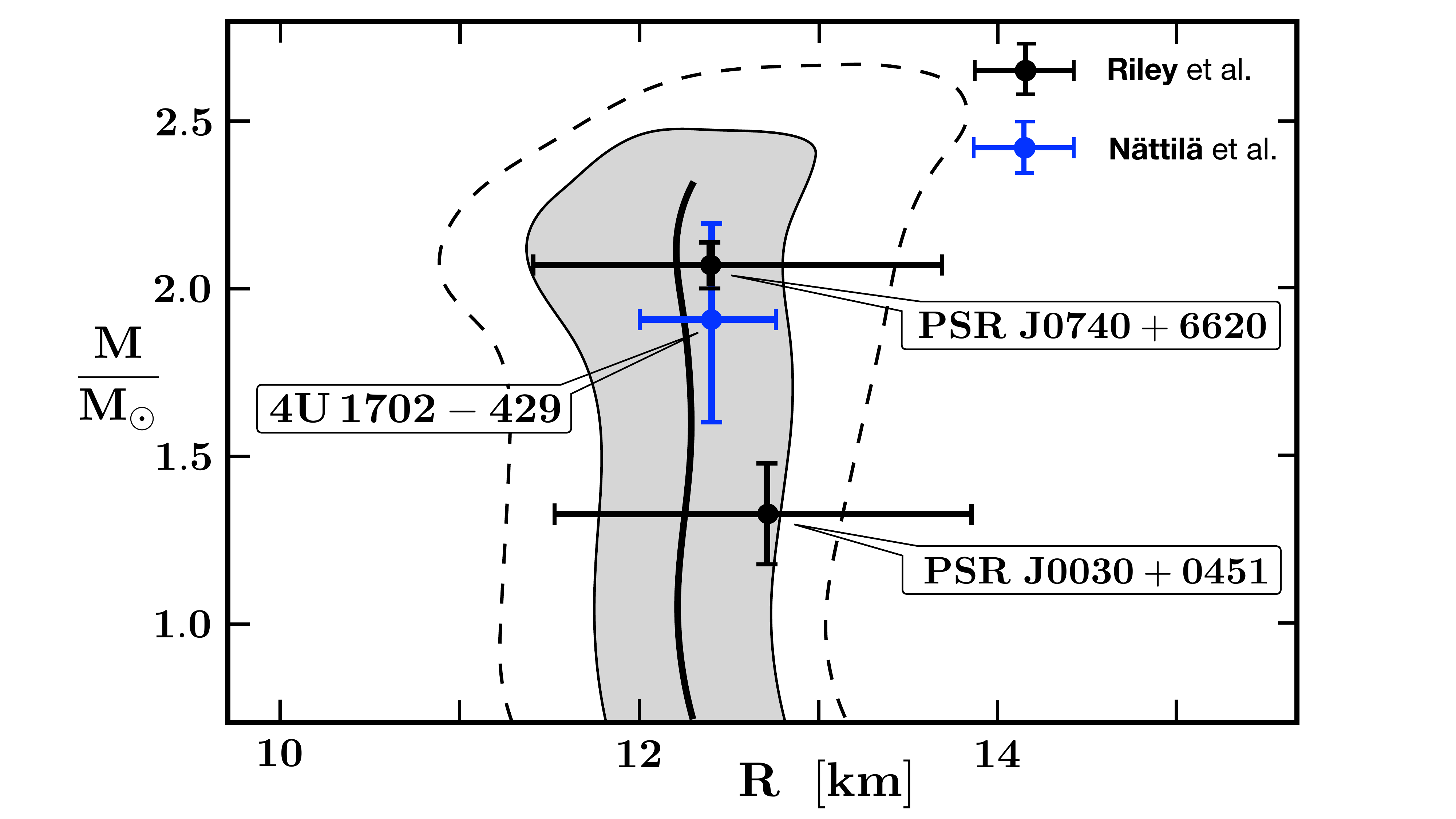}
\caption{Posterior credible bands~\cite{Brandes2023a} 
of the radius R as a function of the neutron star mass M at the 68\% level (gray band) and 95\% level (dashed line)  compared to the analysis of NICER data  by \mbox{Riley et al.} for PSR J0030+0451 and PSR J0740+6620~\cite{Riley2019, Riley2021}.  In addition, the mass--radius credible interval at the 68\%  level of the thermonuclear burster 4U 1702-429 (blue) is displayed~\cite{Nattila2017} (which was not included in the Bayesian analysis).\label{fig5}}     
\end{figure}

Numerically solving the Tolman--Oppenheimer--Volkov (TOV) equations leads to posterior credible bands for the mass--radius relation of neutron stars, as displayed in Figure~\ref{fig5}.  Notably, the median of $R(M)$ suggests an almost constant radius $R$, which is  independent of $M$.

The  density profiles $\rho(r)$ computed for neutron stars in the mass range 1.4--2.3 $M_\odot$ reach central baryon densities,  $\rho_c = \rho(r=0)$,  that are systematically below $5\,\rho_0$ in the cores of even the heaviest stars.  For example,  
\begin{eqnarray}
\rho_c(1.4\,M_\odot) = (2.6\pm 0.4)\,\rho_0\qquad  \text{and}\qquad  \rho_c(2.3\,M_\odot) = (3.8\pm 0.8)\,\rho_0~,
\label{eq:density}
\end{eqnarray}
at the 68\% level~\cite{Brandes2023a}.  In a baryonic picture of neutron stars,  this implies that the average distance between baryons even in the highly compressed star center exceeds 1 fm,  more than twice the characteristic short-range hard-core distance of 1/2 fm in  nucleon--nucleon interactions.  We return to this point at a later stage.

\begin{figure}[H]
\includegraphics[width=8.8cm]{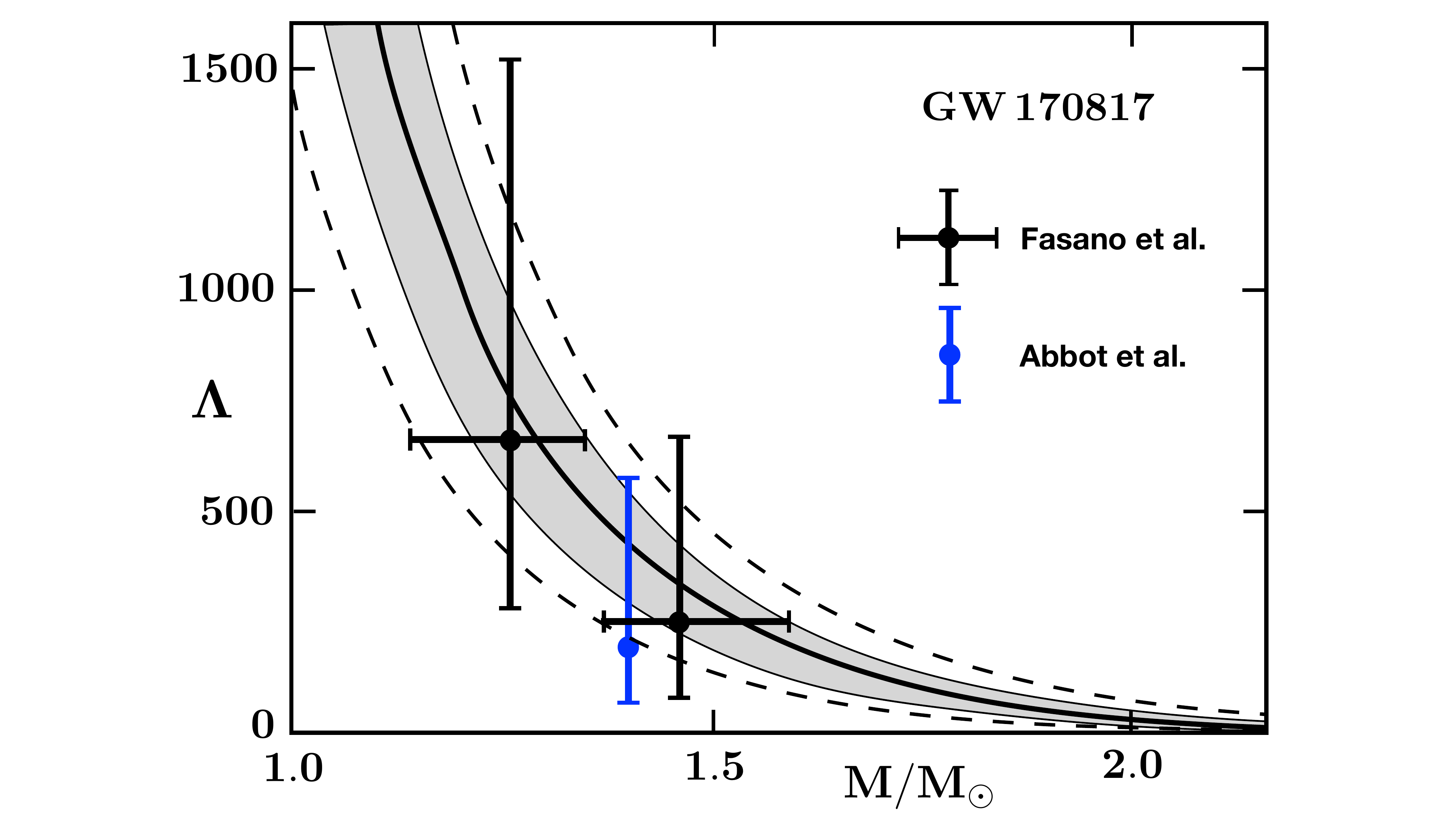}
\caption{Posterior %MDPI: We moved the figure to the end of the first mention, please confirm the revision.
% Author reply: I confirm.
 credible bands~\cite{Brandes2023a} 
of the tidal deformability as a function of neutron star mass  at the 68\% level (gray band) and 95\% level (dashed lines)  compared to values deduced from the merger event GW170817~\cite{Abbott2019,Fasano2019}.  \label{fig6} }     
\end{figure}

For each given EoS,  the TOV equations can be solved in combination with equations for the tidal deformability $\Lambda$.  This leads to posterior probability bands,  $\Lambda(M)$,  as a function of neutron star mass, as shown in Figure~\ref{fig6}.  The large uncertainties of the empirical values deduced from the analysis of the gravitational wave events do  not allow one to constrain the tidal deformability bands to a stronger degree so far.

\section{Constraints on Phase Transitions in Neutron Stars}

The nature and location of the transition from dense nuclear or neutron matter to quark and gluon degrees of freedom is still largely unknown.  Lattice computations have been successful in exploring the QCD phase structure at high temperatures and vanishing baryon chemical potential,  establishing that the transition is a continuous crossover at a temperature of $T_c \simeq 155$ MeV.  However,  at nonzero baryon densities, the notorious sign problem in the Euclidean action prevents a similarly systematic approach.  As a consequence,  studies of the phase diagram at low temperatures and high baryon densities are primarily based on models with varying assumptions about the relevant active degrees of freedom~\cite{Sumiyoshi2023}.

\begin{figure}[H]
\includegraphics[width=8.8cm]{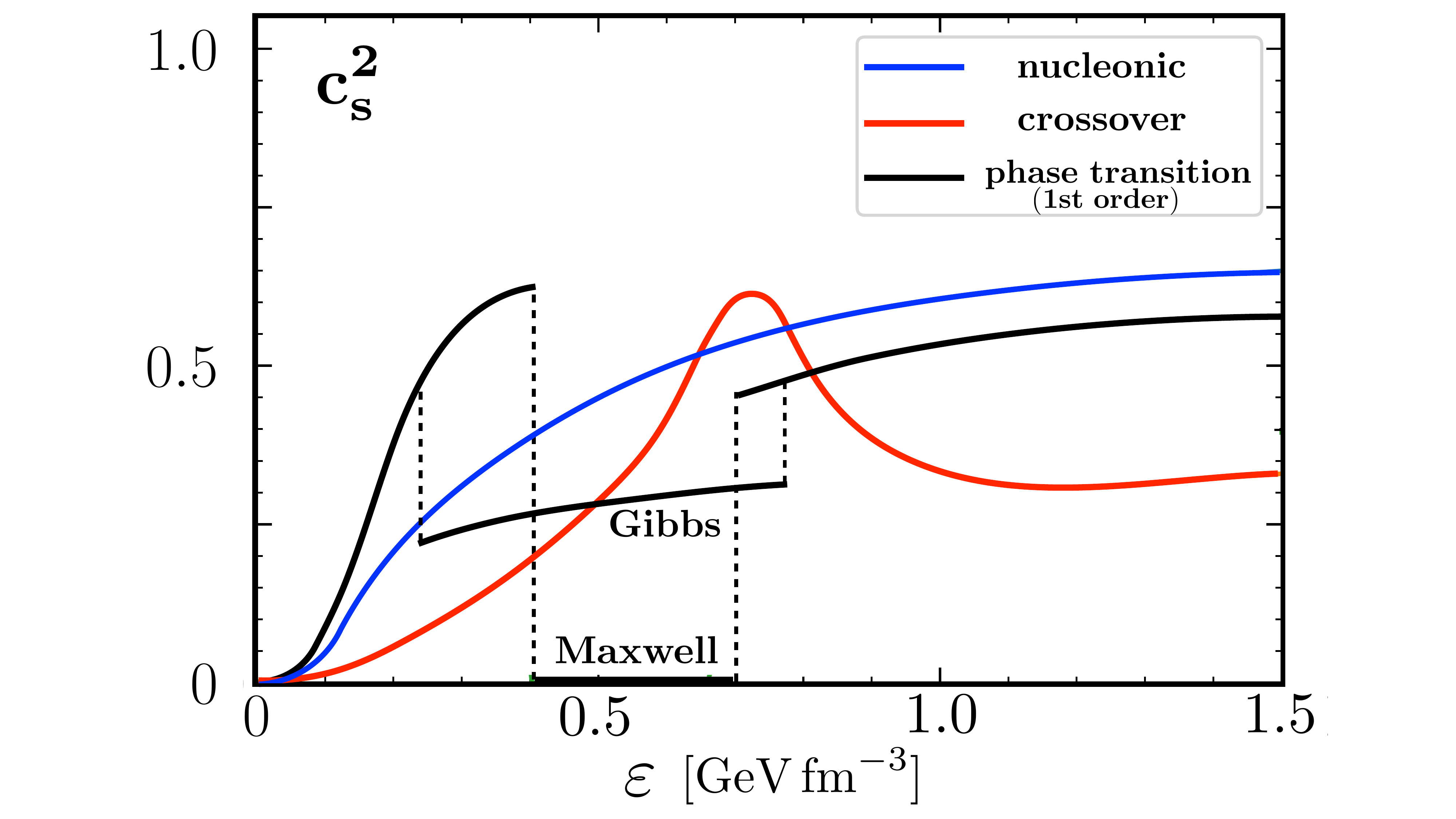}
\caption{Characteristic behaviors of the squared sound velocity in the presence of a phase transition or a crossover.  \label{fig7} }     
\end{figure}

A key element of this discussion is the question about a possible phase transition from spontaneously broken to restored chiral symmetry in QCD.  At high baryon densities beyond $\rho > 2\,\rho_0$, various hypotheses have been explored,  ranging from a chiral first-order phase transition to continuous hadron--quark crossovers~\cite{Fukushima2011,Holt2016, McLerran2019, Baym2019,Fukushima2020,Kojo2022}.  
The behavior of the sound speed $c_s$ is a prime indicator for phase transitions or crossovers.  This is schematically illustrated in Figure~\ref{fig7}, which shows typical patterns of such transitions as they would show up in the sound speed.  For example,  a first-order phase transition with Maxwell construction is characterized by a region of constant pressure over an interval of density (or energy density) in which two phases coexist.  Between the lower and upper endpoints of this phase coexistence interval, the squared sound velocity $c_s^2 = \partial P/\partial\varepsilon$ jumps to zero and back.  In a first-order phase transition with Gibbs construction, the pressure in the mixed phase is not constant, but $\partial P/\partial\varepsilon$ still changes discontinuously~\cite{Han2019}.  A crossover has no such discontinuities but features a pronounced maximum in $c_s^2$.  Purely nucleonic scenarios lead instead to a continuously rising sound speed.

\subsection{Evidence Against a Very Low Squared Sound Speed in Neutron Stars}

One indication of the possible appearance of a first-order phase transition would be a very low squared sound velocity $(c_s^2 \leq 0.1)$ within the range of energy densities relevant for neutron stars.  Let us introduce $c_{s,min}$ as the minimum speed of sound following a maximum at lower densities.  Given the inferred bands of probability distributions for $c_s^2$ in Section~\ref{sec-2.2} and \mbox{Figure \ref{fig1}},  one can quantify the evidence for two competing scenarios: minimum $c^2_{s,min} > 0.1$ versus $c^2_{s,min} \leq 0.1$;  in terms of corresponding Bayes factors,  ${\cal B}^{c^2_{s,min} > 0.1}_{c^2_{s,min} \leq 0.1}$.  A detailed analysis of such Bayes factors was carried out in~\cite{Brandes2023a} for the minimum speed of sound up to different  maximum masses of neutron stars, $M_{max}$.  Bayes factors exceeding 10 or 100 indicate strong or extreme evidence,  respectively,  for the scenario where the minimum $c_s^2$ remains larger than 0.1.  This is the case for all masses $M_{max}\leq 2.1\,M_\odot$, as demonstrated in Table \ref{tab1}.  Only for even heavier neutron stars does this evidence turn out to be lower.  It is worth noting that the inclusion of the heavy black widow pulsar PSR J0952--0607 plays an important role that strengthens the evidence against $c_s^2 \leq 0.1$ further in comparison with previous analyses, which did not include this object~\cite{Brandes2023}.  Squared sound speeds $c_s^2 > 0.1$ in the cores of neutron stars with $M_{max}\leq 2.0\,M_\odot$ were also found in~\cite{Altiparmak2022,Annala2023} without the inclusion of the black widow pulsar.

These limiting bounds on very low speeds of sound suggest that the appearance of a strong first-order phase transition with Maxwell construction is unlikely for most of the known neutron stars.  An additional feature of a Maxwell-constructed first-order transition is an extended phase coexistence region.  This will be examined more closely in the following subsection.

\begin{table}[H]
\caption{Bayes %MDPI: We made the first line bolded as table head, please confirm.
 % Author reply: I confirm.
 % We also changed 'which are indicated' back to 'as indicated'.
 factors ${\cal B}^{c^2_{s,min} > 0.1}_{c^2_{s,min} \leq 0.1}$ comparing EoS samples with competing scenarios: (a) minimum squared speed of sound (following a maximum) with $c^2_{s,min} > 0.1$ versus (b) equations of state with $c^2_{s,min} \leq 0.1$.  The minimum speeds of sound are computed up to the maximum neutron star masses, as indicated  (taken from~\cite{Brandes2023a}). \label{tab1}}
\newcolumntype{C}{>{\centering\arraybackslash}X}
\begin{tabularx}{\textwidth}{CCCCCC}
\toprule
\boldmath{$M_{max}/M_\odot$}	& \textbf{1.9}	 & \textbf{2.0} & \textbf{2.1} & \textbf{2.2} & \textbf{2.3}\\
\midrule
	${\cal B}^{c^2_{s,min} > 0.1}_{c^2_{s,min} \leq 0.1}$ & 500.9 & 229.8 & 15.0 &  3.6 & 2.2\\
\bottomrule
\end{tabularx}
\end{table}

\subsection{Evidence Against a Strong First-Order Phase Transition in the Cores of Neutron Stars}
\label{section:phasetrans}

The relatively moderate baryon densities inferred in the cores of neutron stars together with the evidence against very small sound speeds can be complemented by an additional study that is specific to first-order phase transitions with Maxwell construction.  Such phase transitions are characterized by a domain of phase coexistence that extends over a certain range of baryon densities,  $\Delta\rho$.  The width of this domain,  $\Delta\rho/\rho$ (with $\rho$ the density at which the coexistence interval starts),  is a measure of the strength of the phase transition,  i.e.,~the magnitude of surface tension between the two coexisting phases.  We refer to a `strong' first-order phase transition if the width of the mixed phase is $\Delta\rho/\rho > 1$.  One example is the liquid--gas phase transition in symmetric nuclear matter at temperatures of \mbox{$T < 15$ MeV~\cite{Wellenhofer2014, Brandes2021}}.  In contrast, a `weak' first-order phase transition has a value of $\Delta\rho/\rho$ that is small compared to~unity.

\begin{figure}[H]
\includegraphics[width=8.8cm]{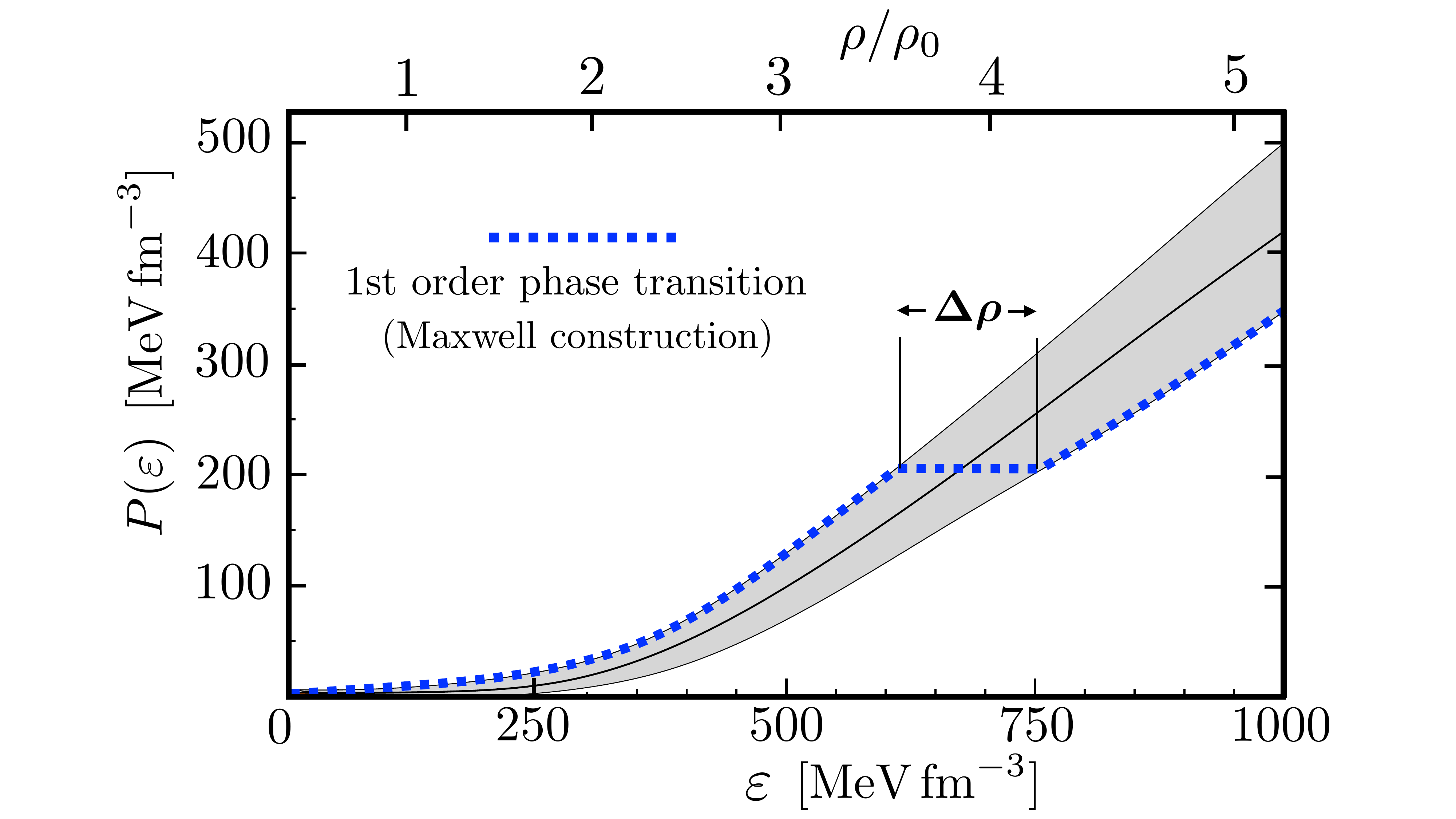}
\caption{Illustration %MDPI: We moved the figure to the end of the first mention, please confirm the revision.
 % Author reply: I confirm.
 of the constraint on the maximum width of a Maxwell-constructed coexistence region for a first-order phase transition within the 68\% credible band of $P(\varepsilon)$.   The %MDPI: Please confirm if the bold is unnecessary and can be removed. The following highlights are the same
 % Author reply: I confirm that the bold can be removed.
 upper (baryon density) scale refers to the median of the $P(\varepsilon)$ distribution, as in Figure~\ref{fig2}.\label{fig8}}     
\end{figure}

The posterior credible bands inferred from neutron star data,  as displayed in Figure~\ref{fig2},  permit a systematic study of the maximum possible phase coexistence widths over the range of relevant energy densities.  Starting from $P(\varepsilon)$, the Gibbs--Duhem relation is used to re-express pressure as a function of baryon density,  $P(\rho)$.  In this way, posterior credible bands for $P(\rho)$ can be derived.  These constrain the maximum possible widths of mixed-phase domains,   $(\Delta\rho/\rho)_{\text{max}}$.  An example is shown in Figure~\ref{fig8}.  In the analysis in~\cite{Brandes2023a}, it turned out that these maximum possible phase coexistence regions were narrow: $(\Delta\rho/\rho)_{\text{max}} \simeq 0.2$ at the 68\% level.  Even within the 95\% credible bands, this width did not exceed 0.3.  In fact, one finds that $(\Delta\rho/\rho)_{\text{max}}$ stays nearly constant as a function of baryon density $\rho$ (taken at the starting point of the mixed-phase interval) over the whole region of densities $\rho \simeq 2 - 5\,\rho_0$ that are relevant for neutron stars.

We note that the study performed in~\cite{Gorda2023} also discussed restrictive constraints on strong first-order phase transitions but did not include the NICER and J0952-0607 black widow pulsar data.  Even with this much more limited database, the authors found possible phase coexistence intervals of only $\Delta\rho/\rho\lesssim 0.5$ for first-order phase transitions within neutron stars.  As can be expected,  the additional radius constraints imposed by the NICER data  constrain the maximum extension of the  phase coexistence region even further. Moreover, we reiterate the additional restrictive impact of J0952-0607 in constraining the possible occurrence of phase transitions. This is at variance with % Please confirm that meaning has been retained 
 % Author reply: Meaning has not been retained, we prefer the expression 'at variance' over 'in disagreement' 
 conclusions drawn about a strong deconfinement transition in neutron star matter~\cite{Annala2020, Annala2023} on the basis of a dataset that did not yet include the heaviest pulsar observed so far.

	\subsection{Intermediate Summary}
The conclusion drawn in our reported analysis is that only a weak first-order phase transition with a mixed-phase width $\Delta\rho/\rho \leq 0.2-0.3$ in a Maxwell construction can be realized inside neutron stars within the posterior credible bands of the inferred equation of state.  A strong first-order phase transition is, thus, very unlikely to occur~\cite{Brandes2023a}.  It is worth pointing out that this conclusion is entirely based on the analysis of observational data,  independent of any specific EoS model.  As will be discussed later,  this rules out some relativistic mean-field models that suggested the occurrence of a strong (chiral) first-order phase transition already at densities below those encountered in neutron star cores.  On the other hand, the empirical uncertainty bands still leave room for a continuous transition, such as a hadron-to-quark crossover~\cite{Baym2019, McLerran2019, Fukushima2020, Kojo2022}. The possibility of a purely baryonic equation of state with a monotonically rising speed of sound is also not excluded~\cite{Brandes2023}.

\section{Phenomenology and Models}
The previous sections have shown how present and forthcoming neutron star data (masses, radii, and tidal deformabilities) provide constraints for the equation of state and for possible phase transitions in cold,  dense baryonic matter.  The enhanced stiffness of the EoS at high baryon densities is a necessary feature; sufficiently high pressures are required to support neutron stars with masses around and beyond $2\,M_\odot$ at relatively large radii (\mbox{$R\simeq$ 11--%MDPI: We revised the minus sign into a en dash. Please confirm.
% Author reply: I confirm.
13 km}).  The central core densities are not as extreme as was previously imagined.  Even in the core of a 2.3 $M_\odot$ neutron star, the baryon density at its center stays below five times the equilibrium density of % Please confirm that meaning has been retained 
% Author reply: The meaning has not been reatined. It should be '... the equilibrium density of ...' instead.
 nuclear matter (at the 68\% credible level;  see Equation~(\ref{eq:density})).  The upper border of the 95\% credible interval of the inferred central density $\rho_c(2.3\,M_\odot)/\rho_0 =3.8^{+1.6}_{-1.3}$~\cite{Brandes2023a} lies slightly above $5\,\rho_0$, but with a very low probability.  This observation has consequences for the interpretation of the possible structure and composition of neutron star matter, which we discuss in the following.

\subsection{Reminder of Low-Energy Nucleon Structure and a Two-Scale Scenario}
\label{section:chiral}

Spontaneously broken chiral symmetry, as the long-wavelength manifestation of QCD, governs the low-energy structure and dynamics of hadrons, including nucleons and pions.  As chiral Nambu--Goldstone bosons,  pions play a distinguished role in this context.  Models based on chiral symmetry often view the nucleon as a complex system of two scales~\cite{TW2001}: a compact hard core that hosts the three valence quarks and, thus, encloses the baryon number  and a surrounding quark--antiquark cloud in which pions figure prominently as the `soft' degrees of freedom. 

Such a two-scale scenario is manifested in empirical form factors and sizes of  nucleons.  Consider, for example, the proton and neutron electromagnetic form factors and their slopes at zero momentum transfer, which determine the corresponding mean square radii.  The empirical r.m.s. %MDPI: Please confirm that meaning has been retained 
% Author reply: The meaning has not been reatined. The correct nomenclature is  the 'r.m.s. proton charge radius'.
 proton charge radius,  $\langle r_p^2\rangle^{1/2} = 0.840\pm 0.003$ fm,  combined with six times the slope of the neutron electric form factor,  $\langle r_n^2\rangle = -0.105\pm 0.006$ fm$^2$,  gives the isoscalar and isovector mean square radii of the nucleon,  $\langle r^2_{S,V}\rangle = \langle r_p^2\rangle \pm \langle r_n^2\rangle$, with the following resulting values~\cite{Lin2022}:

\begin{eqnarray}
\sqrt{\langle r_S^2\rangle} \simeq 0.78 \,\text{fm}~,\quad ~~\sqrt{\langle r_V^2\rangle} \simeq 0.90 \,\text{fm}~.
\end{eqnarray}
Each of the nucleon form factors $G_i(q^2)$ related to a current with index $i$ has a representation in terms of a dispersion relation,  
\begin{eqnarray}
G_i(q^2)=G_i(0)+\frac{q^2}{\pi}\int_{t_0}^\infty dt\, \frac{\text{Im} \,G_i(t)}{t(t-q^2-i\epsilon)}~,
\end{eqnarray}
% We changed 'a squared four-momentum transfer' to 'the sqaured four momentum-transfer'
\textls[-20]{with the squared four-momentum transfer $q^2=q_0^2 - \vec{q}^{\,2}$.
The mean square radii are given as}
\begin{eqnarray}
\langle r_i^2\rangle = \frac{6}{G_i(0)} \frac{dG_i(q^2)}{dq^2}\Big|_{q^2=0}=
%\langle r_i^2\rangle = 6\frac{dG_i(q^2)}{dq^2}\Big|_{q^2=0}= 
\frac{6}{\pi}\int_{t_0}^\infty\frac{dt}{t^2}S_i(t)~,
\end{eqnarray}
where the distribution $S_i(t)= \text{Im}G_i(t)/G_i(0)$ represents the spectrum of intermediate hadronic states through which the external probing field couples to the respective nucleon current.  For example,  the isovector charge radius reflects the interacting two-pion cloud of the nucleon governed by the $\rho$ meson and a low-mass tail extending down to the $\pi\pi$ threshold,  $t_0 = 4m_\pi^2$.  The isoscalar charge radius is related to the three-pion spectrum, which is strongly dominated by the narrow $\omega$ meson~\cite{Kaiser2019} and starts at $t_0 = 9m_\pi^2$.  The isoscalar charge form factor of the nucleon,  $G_S^E(q^2)$ (with $G_S^E(0) =1$),  is particularly suitable for discussing a delineation between the `core' and `cloud' parts of the nucleon~\cite{Brown1986, Meissner1987}. The vector meson dominance principle implies,  in its simplest version,  a representation of the form
\begin{eqnarray}
G_S^E(q^2) = \frac{F_B(q^2)}{1+|q^2|/m_\omega^2}~.
\end{eqnarray}
The form factor $F_B(q^2)$ of the baryon number distribution in the nucleon core acts as a source for the $\omega$ field that propagates with its mass $m_\omega$.  Introducing the mean square radius of the baryon core,  $\langle r_B^2\rangle = 6\frac{dF_B(q^2)}{dq^2}|_{q^2=0}$,  the mean square isoscalar charge radius becomes
\begin{eqnarray}
\langle r_S^2\rangle = \langle r_B^2\rangle + \frac{6}{m_\omega^2}~.
\end{eqnarray}
Using  $m_\omega = 783$ MeV, the estimated baryonic core radius is 
\begin{eqnarray}
\sqrt{\langle r_B^2\rangle} \simeq  0.47\,\text{fm}~.
\end{eqnarray}
A nucleon core size of about 1/2 fm is characteristic of chiral `core + cloud' models.  It also holds up in more detailed treatments of the spectral distributions governing the nucleon form factors~\cite{Lin2022}.  The inclusion of additional $\phi$ meson and $\pi\rho$ continuum contributions  in the spectral function of $G_S^E(q^2)$ moves the core radius to just slightly larger values.

Consider as another example the form factor associated with the axial vector current of the nucleon.  The corresponding mean-square axial radius deduced from neutrino--deuteron scattering data is reported as~\cite{Hill2018} 
\begin{eqnarray}
\langle r_A^2\rangle = (0.46\pm 0.22)\,\text{fm}^2~.
\end{eqnarray}
A schematic axial vector dominance picture % Please confirm that meaning has been retained
% Author reply: Meaning has not been retained, it should be 'A schematic axial vector dominance picture ...'. 
 would assign a dominant part of the `cloud' contribution to this form factor through the spectrum of the $a_1$ meson with its large width.  If an approximate scale of this `cloud' part is identified with $\delta\langle r_A^2\rangle \sim 6/m_a^2$ using the physical $a_1$ mass, $m_a\simeq 1.23$ GeV,  one finds 

\begin{eqnarray}
\sqrt{\langle r_A^2\rangle}_{\text{core}} = [\langle r_A^2\rangle - \delta\langle r_A^2\rangle]^{1/2} 
\simeq 0.55\,\text{fm} 
\end{eqnarray}
with an estimated uncertainty of about 25\%.  

Yet another interesting piece of information is the mass radius % Please confirm that meaning has been retained 
% Author reply: Meaning has not been retained, the correct nomenclature '... the mass radius of the proton ...'
 of the proton deduced from $J/\Psi$ photoproduction data~\cite{Kharzeev2021}.  It involves the form factor of the trace,  $T_\mu^\mu$,  of the nucleon's energy--momentum tensor and is supposed to be dominated by gluon dynamics at the center of the nucleon.  Low-mass (e.g.,  two-pion) components are suppressed.  The result quoted in~\cite{Kharzeev2021},
\begin{eqnarray}
\langle r^2 \rangle_\text{mass}^{1/2} =  (0.55\pm 0.03)\,\text{fm}~,
\end{eqnarray}
is, once again, remarkably close to an assumed `core' size scale of $\sim 1/2$ fm.  

These empirical considerations motivate a picture of the nucleon as containing a compact `hard'  core in which the valence quarks and their baryon number are confined,  and a `soft' surface of quark--antiquark pairs forming a mesonic cloud.  This structure has implications for the behavior of nucleons in dense baryonic matter.  With a core size of $R_\text{core} \sim 1/2$ fm and a cloud range given, e.g.,  by the proton charge radius, $R_\text{cloud} \sim 0.84$ fm, there is a significant separation of volume scales in vacuum: $(R_\text{cloud}/R_\text{core})^3 \sim 5$.  

This scale separation is expected to increase further in dense baryonic matter for the following reasons.
The properties of the soft multi-pion cloud are closely tied to spontaneously broken chiral symmetry and the approximate Nambu--Goldstone boson nature of the pion.  The size of this cloud is expected to increase with density,  along with the decreasing in-medium pion decay constant,  $f_\pi^*(\rho)$,  which acts as a chiral order parameter.  The baryonic core,  on the other hand,  is governed by gluon dynamics,  without a leading connection to chiral symmetry in QCD.  This core is, therefore, expected to be quite stable against changes from increasing density up until the compact hard cores begin to touch and overlap.  
\begin{figure}[H]
\includegraphics[width=9.5cm]{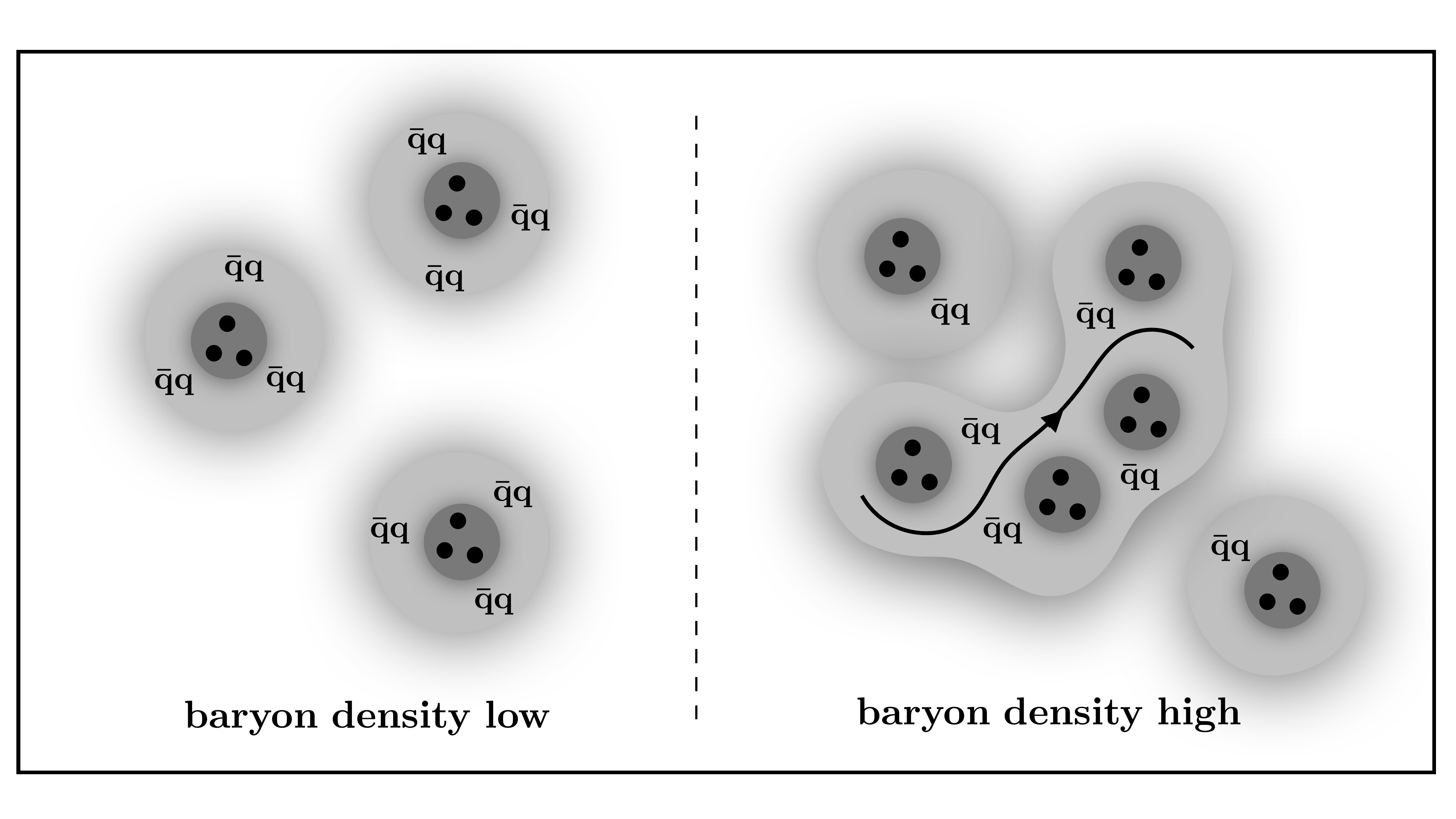}
\caption{Sketch of low- and high-density baryonic matter.  Baryons (e.g.,  nucleons) are viewed as valence quark cores surrounded by clouds of quark--antiquark pairs (e.g.,  chiral meson clouds).  At densities of $\rho \gtrsim 2-3\,\rho_0$, the percolation of quark--antiquark pairs over larger distances starts, as indicated.  Valence quark cores begin to touch and overlap at baryon densities $\rho \gtrsim 5\,\rho_0$.  \label{fig9} }     
\end{figure}
What arises in this way is a two-scale scenario for dense baryonic matter, as described in~\cite{Fukushima2020} and sketched in Figure \ref{fig9}.  At $\rho \simeq \rho_0$, the tails of the meson clouds of nucleon pairs overlap, resulting in two-body exchange forces.  As the average distance between nucleons decreases with increasing density,  around $\rho \gtrsim 2 - 3\,\rho_0$,  the soft clouds of $q\bar{q}$ pairs start to be delocalized.  Their mobility expands over larger distances in a percolation process involving larger numbers of nucleons.  In the terms of conventional nuclear physics,  this corresponds to the emergence of many-body forces,  the strength of which grows with increasing density.  At that stage, the baryonic cores are still separated but subject to increasingly repulsive Pauli principle effects.  The cores begin to touch and overlap at average nucleon--nucleon distances of $d \lesssim 1$ fm,  corresponding to densities $\rho \gtrsim 5\,\rho_0$.  Further compression of baryonic matter would still have to overcome the strongly repulsive hard core in the nucleon--nucleon interaction.  Recalling the inferred credible intervals of the central densities in heavy neutron stars  (Equation~(\ref{eq:density})), one concludes that a phase transition to valence quark deconfinement does not seem likely in a two-scale picture and under the conditions in the cores of neutron stars unless they are extremely heavy.  

An interesting and closely related result emerges from a detailed analysis of $y-$scaling in electron--nucleus scattering at large momentum transfers ($|\vec{q}| \gtrsim 1$ GeV) and low energy transfers~\cite{Benhar2023}.  The persistently observed $y-$scaling under these kinematical conditions implies that the electrons still scatter from strongly correlated pairs of nucleons,  rather than quarks,  at short distances.  The conclusion drawn in~\cite{Benhar2023} is that at local densities as large as five times $\rho_0$,  nuclear matter still appears to behave as a collection of nucleons. 

\subsection{Quark--Hadron Continuity and Crossover}

While a strong first-order phase transition in neutron star matter seems unlikely based on the current empirical observations,  a continuous crossover from hadrons to quarks is still possible within the present data-driven constraints.  Such a scenario is realized, for example, in the QHC21 equation of state~\cite{Kojo2022}.  It features a smooth interpolation between low and high densities from nuclear to quark matter regimes.  The quark matter phase is described by a three-flavor Nambu--Jona-Lasinio (NJL) model that includes pairing (diquark) degrees of freedom and a strongly repulsive vector coupling between quarks.  In order to build up the necessary pressure at high densities,  this vector coupling must be comparable to or larger than the strength of the standard scalar--pseudoscalar interaction in the NJL model.  The density at which the interpolated turnover to quark matter takes place in the QHC21 EoS is chosen as $\rho\gtrsim 3.5\,\rho_0$,  within the range of central densities that can be reached in $M\gtrsim 2\,M_\odot$ neutron stars.  
A crossover from hadrons to quarks could also involve an intermediate phase of quarkyonic matter~\cite{McLerran2019, Fujimoto2023}.

\subsection{Restoration of Chiral Symmetry  in Dense Matter: From First-Order Phase Transition to~Crossover}

The quest for chiral symmetry restoration at high baryon densities---a transition from the spontaneously broken Nambu--Goldstone realization to the unbroken Wigner--Weyl mode---has been a persistent issue for a long time.  A possible first-order chiral phase transition and the existence of a corresponding critical endpoint in the QCD phase diagram have always been topics of prime interest~\cite{Stephanov1998, Fukushima2003, Fischer2014}. Early hypotheses concerning the occurrence of a first-order phase transition were frequently based on Nambu--Jona-Lasinio  (NJL)-type models in mean-field approximation~\cite{Asakawa1989, Klimt1990, Scavenius2001}, which were later extended by \textls[20]{incorporating confinement aspects through added Polyakov-loop degrees of} freedom~\cite{Roessner2007,Roessner2008, Hell2010}. 

The empirical constraints on strong first-order phase transitions in dense neutron star matter, as described in Section \ref{section:phasetrans}, are strikingly at variance with % Please confirm that meaning has been retained 
% Author reply: Meaning has not been retained. We prefer 'variance' over 'disagreement' here.
 previous mean-field (e.g.,~NJL model) predictions. These suggested that a first-order chiral phase transition should already appear at relatively low baryon densities around  $\rho \sim 2-3\,\rho_0$.  A possible explanation for this discrepancy can be found in~\cite{Drews2017, Brandes2021}, where a chiral nucleon--meson (ChNM) field theory was used to explore the effects of fluctuations beyond mean-field (MF) approximation.  The starting point was a relativistic chiral Lagrangian,  ${\cal L}(\Psi; \pi,\sigma;v_\mu)$,  shaped around a linear sigma model with nucleons $(\Psi)$,  pions $(\pi)$, and a scalar $(\sigma)$ field.  Short-range dynamics were parametrized in terms of heavy isoscalar and isovector vector fields $(v_\mu)$. The expectation value $\langle\sigma\rangle$ of the scalar field acted as a chiral order parameter normalized in the vacuum to the pion decay constant, $f_\pi \simeq 92$ MeV.  Two classes of fluctuations beyond MF were then systematically studied: first,  vacuum fluctuations that introduced an additional term proportional to $\sigma^4\ln(\sigma/f_\pi)$ in the MF partition function;  secondly,  fluctuations involving pion loops and nucleon particle--hole excitations.  The vacuum fluctuations can be included in an extended mean-field (EMF) approximation~\cite{Skokov2010}.  Fluctuations involving pion and nucleon loops are computed using non-perturbative functional renormalization group (FRG) methods.  The parameters of the ChNM model---in particular, those related to short-distance dynamics---are fixed to reproduce empirical nuclear physics data~\cite{Drews2017,Brandes2021}.  

\begin{figure}[H]
\includegraphics[width=8.7cm]{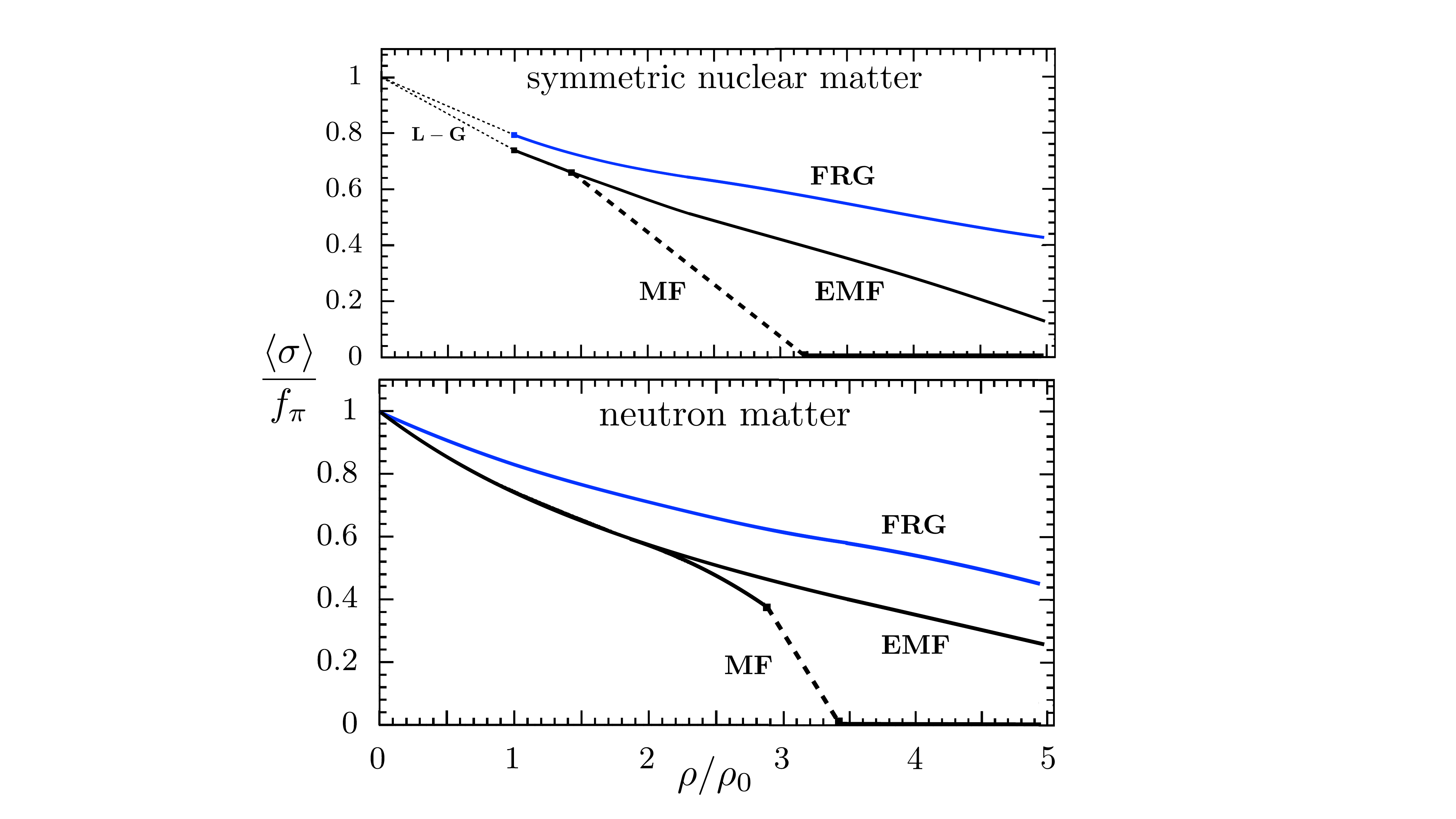}
\caption{\textls[-16]{Chiral order parameters in symmetric nuclear matter and neutron matter at a temperature of $T=0$ as  a function of baryon density in units of nuclear ground state equilibrium density, \mbox{$\rho_0 = 0.16$ fm$^{-3}$}.  Dotted lines: liquid--gas phase transition (L-G) in symmetric nuclear matter.  Dashed lines: first-order chiral phase transitions emerging from the mean-field (MF) approximation of a relativistic chiral nucleon--meson (ChNM) field-theoretical model.  Solid lines show the results of extended mean-field (EMF) calculations (with the inclusion of fermionic vacuum fluctuations)  and functional renormalization group  (FRG) calculations  based on the same ChNM model.  The figures were adapted from Refs.~\cite{Brandes2021, Drews2017}.}  \label{fig10} }     
\end{figure}

Figure~\ref{fig10} demonstrates the important role of fluctuations beyond the mean-field approximation for the chiral order parameter $\langle\sigma\rangle$.  In symmetric nuclear matter, the mean-field approximation of the ChNM model correctly reproduces the first-order liquid--gas phase transition at low density.  However, the chiral order parameter also displays a strong first-order chiral phase transition with a Maxwell-constructed phase coexistence region starting already below $2\,\rho_0$.  For neutron matter,  which has no liquid--gas phase transition,  the MF approximation nevertheless predicts a first-order chiral phase transition at densities around $\rho\sim 3\,\rho_0$, which is well within the range of densities realized in neutron stars.  However,  in both nuclear and neutron matter,  the inclusion of fermionic vacuum fluctuations (i.e.,  the effect of the ground state zero-point energy) in the extended mean-field (EMF) approximation converts the first-order chiral phase transition into a smooth crossover and shifts it to densities  of $\rho > 5\,\rho_0$.  This effect is further enhanced by the additional fluctuations included in the full FRG calculation, as demonstrated in Figure~\ref{fig10}.  As a result, the restoration of  chiral symmetry  is relegated to very high baryon densities beyond the inferred core densities in even the heaviest neutron stars (see Equation~(\ref{eq:density})). 
\textls[-19]{A comparable impact of fluctuations on the phase structure is seen in alternative chiral models~\cite{Gupta2012,Zacci2018}.}

Another approach based on chiral symmetry is the parity-doublet model.  In this model, the active coupled baryonic degrees of freedom are the nucleon with spin-parity $1/2^+$ and its chiral partner with spin-parity $1/2^-$,  where the latter is  identified with the $N^*(1535)$ resonance.  Spontaneous chiral symmetry breaking in vacuum manifests itself in the mass splitting of these two states,  while the in-medium restoration pattern of this symmetry is signaled by the N and N* masses becoming degenerate.  A recent detailed analysis~\cite{Eser2023} of the chiral order parameter in this model using extended mean-field approximation found a chiral phase transition in nuclear matter,  but at extremely high densities $(\rho > 10\,\rho_0)$ that are  far beyond the density scales reached in neutron stars.

\subsection{Dense Baryonic Matter: A Fermi Liquid Picture}

\begin{figure}[H]
\includegraphics[width=10cm]{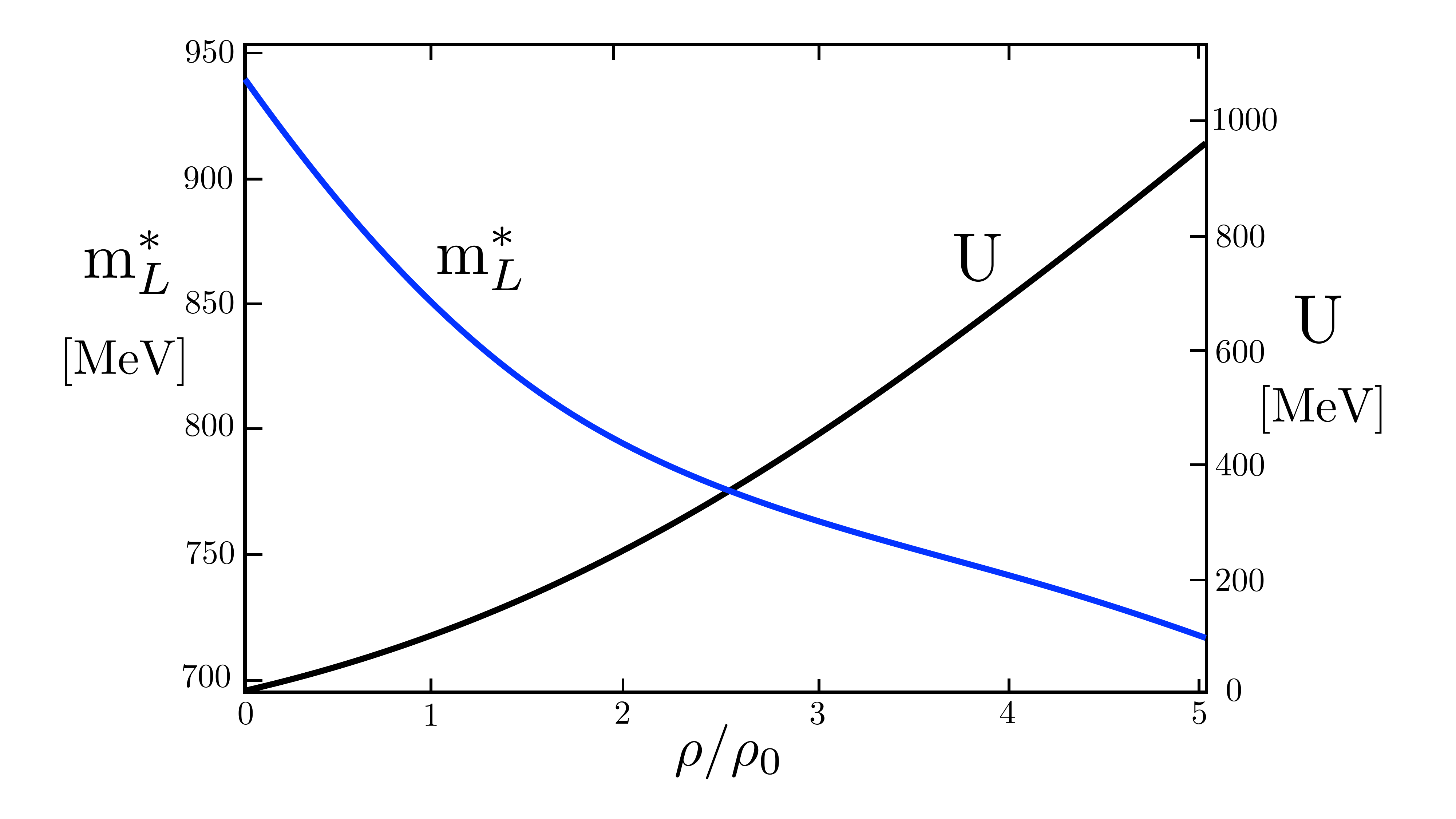}
\caption{The Landau effective mass $m_L^*(\rho) = \sqrt{p_F^2+m^2(\rho)}$ and potential $U(\rho)$ of  quasiparticles representing the median of the posterior distribution of the baryon chemical potential $\mu(\rho)$ (see Figure~\ref{fig3}).  \label{fig11}}     
\end{figure}

The constraints on the equation of state inferred from the current empirical data still permit an interpretation of neutron star core matter in terms of baryonic degrees of freedom,  such as a system dominated by neutrons~\cite{APR1998} with small fractions of protons and perhaps hyperons~\cite{Lonardoni2015, Gerstung2020, Leong2023}.  The inferred baryon chemical potential $\mu(\rho)= \frac{\partial\varepsilon(\rho)}{\partial\rho}$ shown in Figure~\ref{fig3} does not distinguish between different species of baryons.  Its behavior nonetheless displays  increasingly strong correlations at high densities.  It is instructive to analyze the gross properties of this state of matter using the Landau theory of relativistic Fermi liquids~\cite{Baym1976}.  Here, we perform a schematic study assuming ``neutron-like'' quasiparticles~\cite{Friman2019} with a baryon number of $B=1$ and a density-dependent mass of $m(\rho)$ while ignoring other small admixtures in the composition of the dense medium.  These quasiparticles are characterized by their (relativistic) Landau effective mass $m_L^*$ at a Fermi momentum of $p_F = (3\pi^2\,\rho)^{1/3}$, 
\begin{eqnarray}
m_L^*(\rho) = \sqrt{p_F^2 + m^2(\rho)}~,
\label{eq:Landaumass}
\end{eqnarray}
together with an effective potential,  $U(\rho)$,  so that the baryon chemical potential can be written as~\cite{Friman2019}
\begin{eqnarray}
\mu(\rho) = m_L^*(\rho) + U(\rho)~.
\label{eq:chempot}
\end{eqnarray}
The median of the $\mu(\rho)$ posterior credible bands in Figure~\ref{fig3} is now taken as a guiding starting point to extract the baryonic quasiparticle properties.  With an educated ansatz for $m(\rho)$,  the density dependence of the potential $U(\rho)$ can then be deduced and further discussed.  One possible choice is to take $m(\rho)$
(with $m(0) = 939$ MeV) from the non-perturbative FRG calculation employing the chiral nucleon--meson field-theoretical model~\cite{Drews2017} discussed in Section \ref{section:chiral}. The resulting Landau effective mass $m_L^*(\rho)$ is shown in Figure~\ref{fig11} together with the potential $U(\rho) = \mu(\rho) - m_L^*(\rho)$.  The 95\% credible band of $\mu(\rho)$ in Figure~\ref{fig3} leads to an uncertainty of about 15\% for $U$ at high densities.  It is instructive to fit the resulting quasiparticle potential by a series in powers of baryon density for $\rho\lesssim 5\,\rho_0$ (with $\rho_0 = 0.16\,\text{fm}^{-3}$, as before):
\begin{eqnarray}
U(\rho) = \sum_n u_n\left(\frac{\rho}{\rho_0}\right)^n~.
\label{eq:potential}
\end{eqnarray}
The coefficients fitted to the median of $\mu(\rho)$ are 
\begin{equation}
u_1 = 90.9\,\text{MeV},\quad~u_2=15.3\,\text{MeV},\quad~u_3=3.2\,\text{MeV},\quad~u_4=-0.4\,\text{MeV}~.
\label{eq:potential}
\end{equation}
This pattern reflects a hierarchy of many-body correlations,  recalling that the term linear in density represents two-body interactions,  the term of order $\rho^2$ corresponds to short-range three-body forces,  and so forth. The role of the repulsive $N$-body terms with $N>2$ is quite significant; at $\rho\simeq 4\,\rho_0$, corresponding to an average distance of about 1 fm between the baryonic quasiparticles, these terms contribute as much as the two-body forces to \textls[-4]{the potential $U$ and generate the strong pressure to support heavy neutron stars.  Of course, these} statements rely on the ansatz for the density-dependent mass $m(\rho)$, which is guided by the FRG calculations of pure neutron matter.  A small fraction of protons in beta-equilibrated matter will not substantially change this picture.  However,  neutron star core compositions that  qualitatively deviate from this simplified picture may lead to different conclusions.

Finally,  consider the dimensionless Fermi liquid parameters, $F_0$ and $F_1$,  of the spin-averaged quasiparticle interaction.  In terms of the quasiparticle mass and potential,  they are given as~\cite{Friman2019}
\begin{equation}
F_0(\rho)= \frac{p_F}{\pi^2}\left[m(\rho)\frac{\partial m}{\partial\rho}+m_L^*(\rho)\frac{\partial U}{\partial\rho}\right]~,\quad\quad~F_1(\rho) =-\frac{3U(\rho)}{\mu(\rho)}~.
\end{equation}
Further useful relations are $1+F_0 = N(0)(\partial\mu/\partial\rho)$ with the density of quasiparticle states at the Fermi surface,  $N(0) = m_L^*p_F/\pi^2$,  and $1+F_1/3 = 1-U/\mu = m_L^*/\mu$. 

\begin{figure}[H]
\includegraphics[width=9cm]{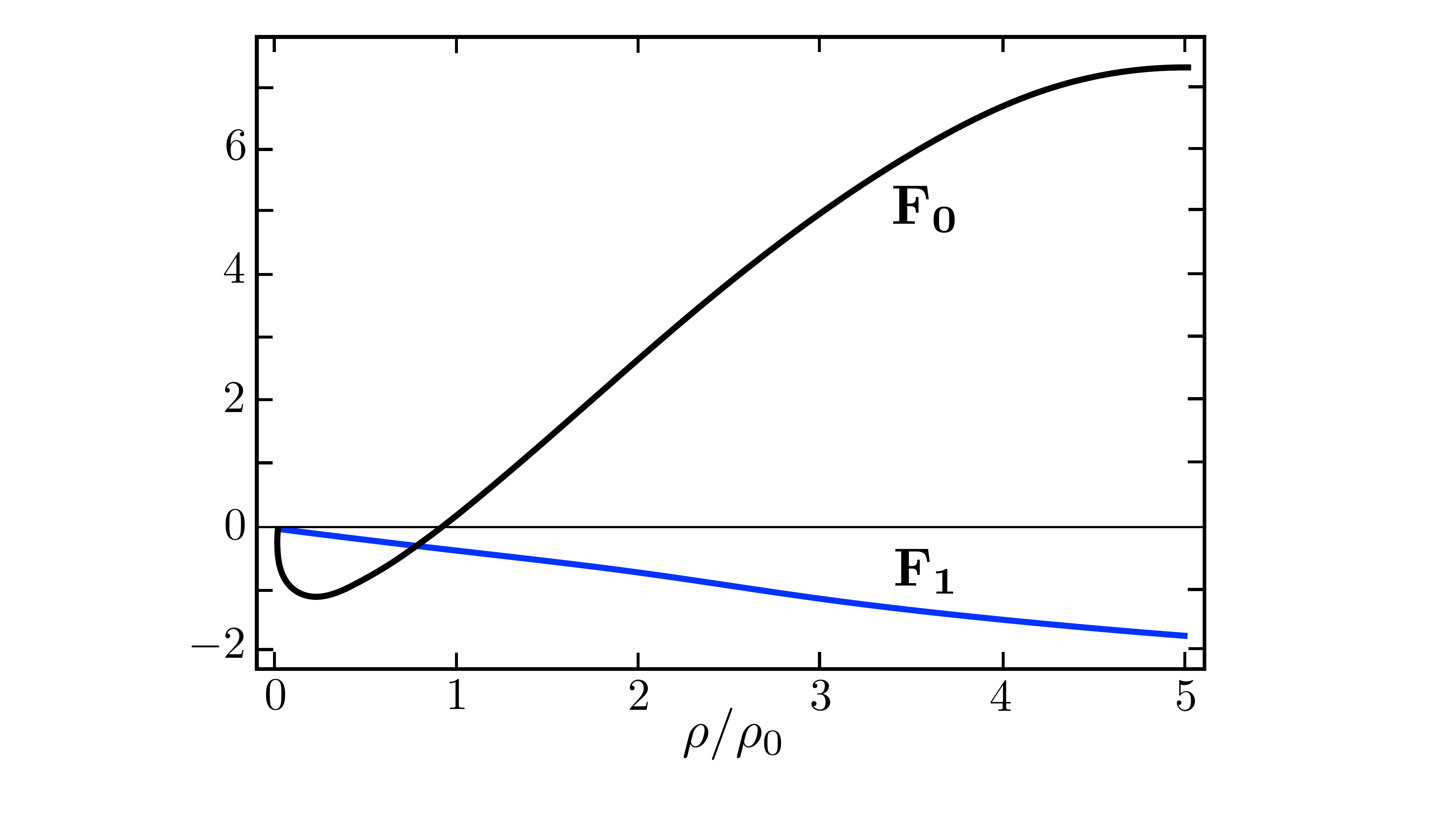}
\caption{The Landau Fermi liquid parameters derived from  quasiparticle properties that are based on the median of the data-inferred baryon chemical potential $\mu$ (see Figure~\ref{fig3}). \label{fig12}}     
\end{figure}

The result in Figure~\ref{fig12} shows a strongly increasing $F_0$ at high densities.  This reflects once again the growing importance of many-body correlations as matter becomes more and more compact.  Such repulsive correlations are responsible for the increase in the sound velocity beyond its canonical conformal limit,  $c_s > 1/\sqrt{3}$,  as seen in Figure~\ref{fig1}.  The Fermi liquid parameter $F_1$ is smaller in magnitude and has a negative slope, indicating the decreasing effective mass at higher densities.  
The Landau parameter $F_0$ displays the typical behavior of a strongly correlated Fermi system.  However, it is interesting  to observe that, in comparison with the leading Landau parameters in liquid $^3$He,  the correlations in neutron star core matter are not extraordinarily strong.  Values of $F_0\approx 9.3$ were reported for $^3$He at zero pressure,  and $F_0\approx 68.2$ at a pressure of 27 bar~\cite{Baym1991}. Accordingly, the Fermi liquid parameters for matter in the density range that is reached inside even the heaviest neutron stars, $\rho\lesssim 5\,\rho_0$, are significantly smaller. From this perspective, the dense baryonic medium encountered in the center of neutron stars is perhaps not as extreme as often imagined.

\section{Concluding Remarks}

The neutron star database, which was recently extended by adding the heaviest pulsar (PSR J0952--0607) observed so far,  requires a much stiffer  equation of state for neutron star matter compared to those of previous analyses. As a consequence, the baryon densities reached in the cores of even very heavy neutron stars do not exceed about five times the density of normal nuclear matter.  In %MDPI: Please confirm if the bold is unnecessary and can be removed. The following highlights are the same.
% Author reply: The bold can be removed.
% Also, we have change 'Chiral models' descriptions' in the next sentence back to 'chiral model descriptions' in the next sentence which is the correct nomenclature!
 a baryon-dominated system, such densities correspond to average distances between any two baryons of 1 fm or larger. Chiral model descriptions of the nucleon suggest a scale separation between a compact valence quark core and a mesonic cloud,  so the nucleonic cores with radii of about 1/2 fm would begin to touch and overlap only in the deep interior of extremely heavy neutron stars.  The appearance of a strong first-order phase transition becomes unlikely under such conditions.  This is consistent with the empirical results, which suggest that extended phase coexistence regions and small minimum sound speeds are not favored by the current data.  This interpretation is further endorsed by the fact that fluctuations beyond the mean-field approximation  convert a possible first-order chiral phase transition into a smooth crossover at high densities.  However, this is at variance %MDPI: Please confirm that meaning has been retained 
 % Author reply: Meaning has not been retained. We prefer 'variance' over 'disagreement' here.
 with conclusions about a strong deconfinement phase transition that have been drawn from datasets that do not yet include the 2.35 solar mass black widow pulsar.  On the other hand, a continuous hadron-to-quark crossover scenario---or alternatively,  baryonic matter as a strongly correlated relativistic Fermi liquid---remains possible within the inferred data-driven  constraints.

\vspace{6pt}

\authorcontributions{Both authors contributed equally to this research article. Both have read and agreed to the published version of the manuscript. %MDPI: For research articles with several authors, a short paragraph specifying their individual contributions must be provided. The following statements should be used ``Conceptualization, X.X. and Y.Y.; methodology, X.X.; software, X.X.; validation, X.X., Y.Y. and Z.Z.; formal analysis, X.X.; investigation, X.X.; resources, X.X.; data curation, X.X.; writing---original draft preparation, X.X.; writing---review and editing, X.X.; visualization, X.X.; supervision, X.X.; project administration, X.X.; funding acquisition, Y.Y. All authors have read and agreed to the published version of the manuscript.'', please turn to the \href{http://img.mdpi.org/data/contributor-role-instruction.pdf}{CRediT taxonomy} for the term explanation. Authorship must be limited to those who have contributed substantially to the work~reported.
% Author reply: I think the above applies to mainly to research articles with many authors, some of which contributing only very little. In our case both authors have contributed equally which I have noew specified.  
}

\funding{This research was partially supported by Deutsche Forschungsgemeinschaft (DFG grant TRR110) and National Natural Science Foundation of China (NSFC grant no. 11621131001) through the Sino-German CRC110 ``Symmetries and the Emergence of Structure in QCD''  and by the DFG Excellence Cluster ORIGINS.}

\dataavailability{The data presented in this study are available on request from the corresponding author. %MDPI: If data are actually included, please use "Data are contained within the article." or "Data are contained within the article and supplementary materials." or "The data presented in this study are available on request from the corresponding author." or "The data presented in this study are not available due to privacy." rather than "Not applicable." in this part. If data are not included, please use "No new data were created or analyzed in this study. Data sharing is not applicable to this article." Please refer to the complete guideline at https://www.mdpi.com/ethics#_bookmark21.
% Author reply: I have specified the data availability.
}

\acknowledgments{Stimulating discussions and communications with Jean-Paul Blaizot,  Kenji Fukushima,  Norbert Kaiser, and Mannque Rho are gratefully acknowledged.  One of the authors (L.B.) thanks the Japan Society for the Promotion of Science and The University of Tokyo for their support and hospitality during his International Fellowship for Research in Japan in the summer of 2023.}

\conflictsofinterest{{The authors declare no conflicts of interest.} }

%\begin{adjustwidth}{-\extralength}{0cm}

\end{paracol}

\reftitle{References}

%\PublishersNote{}
%\end{adjustwidth}
\end{document}